# *Ab initio* derivation of multipolar expansion of optical force

Yikun Jiang,[1,2]* Zhifang Lin* and Jack Ng[3]†

[1]*State Key Laboratory of Surface Physics and Department of Physics,
Fudan University, Shanghai, China*

[2]*Key Laboratory of Micro and Nano Photonic Structures (Ministry of Education),
Fudan University, Shanghai, China*

[3]*Department of Physics and Institute of Computational and Theoretical Studies,
Hong Kong Baptist University, Hong Kong*

*These authors contributed equally to this work.

†jacktfng@hkbu.edu.hk

**Abstract**

Like many other physical quantities, the optical force can be expanded using multipole expansion, which has been done in [Nat. Photon. **5**, 531], up to electric octupole order. However, in that study, the existence of radiation multipoles were pre-assumed, and the role of the fundamental building units, charges, are not evident. Here, we derive the same multipolar expression of optical force by treating the particles as a collection of point charges or point dipoles, which results in more transparent physics and mathematics.

Since the pioneering work of Arthur Ashkin [1,2], optical micromanipulation has developed into a subject of considerable importance wherever small particles are involved [3]. In a typical optical trapping experiment in water, the relevant forces are optical force, Brownian fluctuation, and gravity. With increasing particle size, optical force generally increases but not as fast as gravity does, while the Brownian fluctuation is independent of size according to equipartition theorem. Consequently, optical trapping of nanoparticle is difficult due to Brownian fluctuation, while for macroscopic particles, optical force cannot compete with the gravity. The only regime where optical force is strongly relevant is the Mie regime, where the particle size is comparable with the wavelength, and optical force dominates over the gravity and Brownian fluctuation.

Here, we present an explicit *ab initio* derivation of the analytic multipole expansion of optical force up to the electric octupole order, where the particle is assumed to be consisted of a collection of point charges. Such analytic multipolar expression works for a small to moderately sized Mie particle (up to roughly half a wavelength) and has been presented before [4]. A simpler dipole approximation was also presented previously [5,6]. Compare to previous approaches, our *ab initio* derivation is more fundamental, and it highlights the role of charges in electrodynamics.

Consider a particle consists of a collection of charges illuminated by incident fields $\mathbf{E}$ and $\mathbf{B}$. The charges are assumed to be bounded together by other internal forces that are strong compare to the optical force. Let $\mathbf{r}$ be the particle's center of mass. The time averaged optical force, here after referred simply as the optical force, acting on a collection of charges is generally given by

$$\mathbf{F} = \iiint_V \rho(\mathbf{r})\left[\mathbf{E}_{tot}(\mathbf{r}) + \mathbf{v} \times \mathbf{B}_{tot}(\mathbf{r})\right] dV \quad , \tag{1.1}$$

where $\rho$ is the volume charge density, $\mathbf{v}$ is the velocity field, $\mathbf{E}_{tot}$ and $\mathbf{B}_{tot}$ are the total electric and magnetic fields, respectively. As $\mathbf{E}_{tot}$ are $\mathbf{B}_{tot}$ have two components (the incident field and the field due to charges), the force can also be divided into two components. Assuming that the charge distribution is consisted of point charges, the charge-incident field interaction gives the extinction force

$$\mathbf{F} = \sum_i q_i \left[\mathbf{E}(\mathbf{r}+\mathbf{r}'_i) + \mathbf{v}_i \times \mathbf{B}(\mathbf{r}+\mathbf{r}'_i)\right], \tag{1.2}$$

where $\mathbf{E}$ and $\mathbf{B}$ are the incident field, $q_i$ is the *i*-th point charge, $\mathbf{v}_i$ is the velocity of the *i*-th point charge, $\mathbf{r}'_i$ is the position of the *i*-th charge relative to the center of mass. The charge-charge interaction gives the recoil force [4], which has the form

$$-\frac{k^3 \omega \mu_0}{12\pi} \operatorname{Re}\{\mathbf{p}\times\mathbf{m}*\} - \frac{ck^5 Z_0}{120\pi} \operatorname{Im}\{\ddot{\mathbf{Q}}\cdot\mathbf{P}*\} \tag{1.3}$$

for the leading multipoles. The recoil force is simplest to be derived from a far field integration of the Maxwell stress tensor [4], here we focus on the extinction force.

We consider a monochromatic incident wave with angular frequency $\omega$, and each charge is vibrating about its equilibrium position. We have $\mathbf{r}'_i = \mathbf{r}_{i0} + \mathbf{r}_{ik}$, where $\mathbf{r}'_i$ is the displacement of the charge measured from $\mathbf{r}$, $\mathbf{r}_{i0}$ is the equilibrium position of the charge measured from $\mathbf{r}$, and $\mathbf{r}_{ik}$ is the displacement of the particle measured from $\mathbf{r}+\mathbf{r}_0$. The particle velocity can be written as

$$\mathbf{v}_i = \dot{\mathbf{r}}'_i = \dot{\mathbf{r}}_{ik} = -i\omega\dot{\mathbf{r}}_{ik} \quad , \tag{1.4}$$

where a dot above a variable indicates a time derivative. In the following, since we will eventually take time-averaging, all terms of the form $\dot{\mathbf{F}}$ and $\mathbf{const}\cdot\mathbf{B}$ are treated as zero for convenience, as they will be zero after taking time averaging anyway.

We now expand (1.2) in a Taylor series in $\mathbf{r}'_i$. The zeroth order of (1.2) gives

$$(\sum q_i)\mathbf{E}(\mathbf{r}) = \mathbf{0}, \tag{1.5}$$

where $\sum q_i = 0$ for neutral particle has been used.

To the first order, one has:

$$\begin{aligned}
&\sum_i q_i\{(\mathbf{r}'\cdot\nabla)\mathbf{E} + \mathbf{v}_i \times \mathbf{B}\} \\
&= \sum_i q_i\{(\mathbf{r}'\cdot\nabla)\mathbf{E} - \mathbf{r}' \times \dot{\mathbf{B}}\} \\
&= \sum_i q_i\{(\mathbf{r}'\cdot\nabla)\mathbf{E} + \mathbf{r}' \times (\nabla \times \mathbf{E})\} \\
&= \sum_i q_i\{(\mathbf{r}'\cdot\nabla)\mathbf{E} + \nabla(\mathbf{r}'\cdot\mathbf{E}) - (\mathbf{r}'\cdot\nabla)\mathbf{E}\} \\
&= \nabla(\mathbf{p}\cdot\mathbf{E})
\end{aligned} \tag{1.6}$$

where $\mathbf{p} = \sum_i q_i \mathbf{r}_i' = \int \rho \mathbf{r}' d\mathbf{r}'$ is the dipole moment. Time averaging of (1.6) gives

$$\mathbf{F}_{e-dipole} = \frac{1}{2}\nabla \operatorname{Re}[\mathbf{p}\cdot\mathbf{E}^*(\mathbf{r})], \tag{1.7}$$

which is the optical force for an electric dipole.

To the second order:

$$\begin{aligned}
&\sum_i q_i\left\{\frac{1}{2}\mathbf{r}_i'\mathbf{r}_i' : \nabla\nabla\mathbf{E} + \mathbf{v}_i \times [(\mathbf{r}_i'\cdot\nabla)\mathbf{B}]\right\} \\
&= \sum_i q_i\left\{\frac{1}{2}\mathbf{r}_i'\mathbf{r}_i' : \nabla\nabla\mathbf{E} - \mathbf{r}_i' \times [(\mathbf{v}_i\cdot\nabla)\mathbf{B} + (\mathbf{r}_i'\cdot\nabla)\dot{\mathbf{B}}]\right\} \\
&= \sum q_i\{\frac{1}{2}\mathbf{r}_i'\mathbf{r}_i' : \nabla\nabla\mathbf{E} - \mathbf{r}_i' \times (\mathbf{v}_i\cdot\nabla)\mathbf{B} + (\mathbf{r}_i'\cdot\nabla)[\mathbf{r}_i' \times (\nabla \times \mathbf{E})]\} \\
&= \sum q_i\{\frac{1}{2}\mathbf{r}_i'\mathbf{r}_i' : \nabla\nabla\mathbf{E} - \mathbf{r}_i' \times (\mathbf{v}_i\cdot\nabla)\mathbf{B} + (\mathbf{r}_i'\cdot\nabla)[\nabla(\mathbf{r}_i'\cdot\mathbf{E}) - (\mathbf{r}_i'\cdot\nabla)\mathbf{E}]\} \\
&= \sum q_i\{\frac{1}{2}\mathbf{r}_i'\mathbf{r}_i' : \nabla\nabla\mathbf{E} - \mathbf{r}_i' \times (\mathbf{v}_i\cdot\nabla)\mathbf{B} + [\nabla(\mathbf{r}_i'\mathbf{r}_i' : \nabla\mathbf{E}) - \mathbf{r}_i'\mathbf{r}_i' : \nabla\nabla\mathbf{E}]\} \\
&= \sum q_i\{\nabla(\mathbf{r}_i'\mathbf{r}_i' : \nabla\mathbf{E}) - \frac{1}{2}\mathbf{r}_i'\mathbf{r}_i' : \nabla\nabla\mathbf{E} - \mathbf{r}_i' \times (\mathbf{v}_i\cdot\nabla)\mathbf{B}\} \\
&= \sum q_i\{\frac{1}{2}\nabla(\mathbf{r}_i'\mathbf{r}_i' : \nabla\mathbf{E}) + \frac{1}{2}\nabla(\mathbf{r}_i'\mathbf{r}_i' : \nabla\mathbf{E}) - \frac{1}{2}\mathbf{r}_i'\mathbf{r}_i' : \nabla\nabla\mathbf{E} - \mathbf{r}_i' \times (\mathbf{v}_i\cdot\nabla)\mathbf{B}\}.
\end{aligned} \tag{1.8}$$

One can write

$$\mathbf{F} = \frac{1}{6}\nabla[\vec{\mathbf{Q}} : \nabla\mathbf{E}] + \sum q\{\frac{1}{2}\nabla(\mathbf{r}'\mathbf{r}' : \nabla\mathbf{E}) - \frac{1}{2}\mathbf{r}'\mathbf{r}' : \nabla\nabla\mathbf{E} - \mathbf{r}' \times (\mathbf{v}\cdot\nabla)\mathbf{B}\} \tag{1.9}$$

where

$$\vec{Q} = \sum_i q_i (3\mathbf{r_i'}\mathbf{r_i'} - r_i'^2 \mathbf{I}) \tag{1.10}$$

is the electric quadrupole moment. When defining (1.10), we have used the identity $\mathbf{I}:\nabla \mathbf{E} = \nabla \cdot \mathbf{E} = 0$. Such definition has the advantage that $\vec{Q}$ is traceless and is fully consistent with

$$\vec{Q} = \sum_i q_i (3\mathbf{r_i'}\mathbf{r_i'} - r_i'^2 \mathbf{I}) = \int \rho (3\mathbf{r'}\mathbf{r'} - r'^2 \mathbf{I})d\mathbf{r'} \tag{1.11}$$

where $\rho = q\delta(\mathbf{r} - \mathbf{r_0})$ and $\mathbf{r_0}$ is the position vector of the charge.

The time average of the first term in (1.9) gives

$$\mathbf{F}_{e-quadrupole} = \frac{\nabla}{12} \text{Re}[\ddot{\mathbf{Q}}:\nabla \mathbf{E}^*(\mathbf{r})], \tag{1.12}$$

which is the optical force for an electric quadrupole.

Using the Maxwell equation

$$\nabla \times \mathbf{E} = -\frac{\partial \mathbf{B}}{\partial t}, \tag{1.13}$$

or equivalently its component form $\nabla_i E_j = \nabla_j E_i - \varepsilon_{ijk} \dot{B}_k$, one arrives at

$$\mathbf{r'}\mathbf{r'}:\nabla \nabla \mathbf{E} = r_i' r_j' \nabla_i \nabla_j E_k \hat{e}_k = r_i' r_j' \nabla_i (\nabla_k E_j - \varepsilon_{jkl} \dot{B}_l)\hat{e}_k$$
$$= \nabla[\mathbf{r'}\mathbf{r'}:\nabla \mathbf{E}(\mathbf{r})] - \varepsilon_{jkl} r_i' r_j' \nabla_i \dot{B}_l \hat{e}_k = \nabla[\mathbf{r'}\mathbf{r'}:\nabla \mathbf{E}(\mathbf{r})] + \varepsilon_{jkl}\dot{r}_i' r_j' \nabla_i B_l + \varepsilon_{jkl} r_i' \dot{r}_j' \nabla_i B_l)\hat{e}_k$$
$$= \nabla[\mathbf{r'}\mathbf{r'}:\nabla \mathbf{E}(\mathbf{r})] - (\mathbf{v} \cdot \nabla)(\mathbf{r'} \times \mathbf{B}) - (\mathbf{r'} \cdot \nabla)(\mathbf{v} \times \mathbf{B})$$
$$\tag{1.14}$$

So (1.9) becomes

$$\sum q\{\frac{1}{2}(\mathbf{v} \cdot \nabla)(\mathbf{r'} \times \mathbf{B}) + \frac{1}{2}(\mathbf{r'} \cdot \nabla)(\mathbf{v} \times \mathbf{B}) - \mathbf{r'} \times (\mathbf{v} \cdot \nabla)\mathbf{B}\}$$
$$= \sum \frac{1}{2}q\{(\mathbf{r'} \cdot \nabla)(\mathbf{v} \times \mathbf{B}) - (\mathbf{v} \cdot \nabla)(\mathbf{r'} \times \mathbf{B})\}$$
$$= \sum \frac{1}{2}q\{\nabla[\mathbf{r'} \cdot (\mathbf{v} \times \mathbf{B})] - \mathbf{r'} \times \nabla \times (\mathbf{v} \times \mathbf{B}) - (\mathbf{v} \cdot \nabla)(\mathbf{r'} \times \mathbf{B})\} \tag{1.15}$$
$$= \sum \frac{1}{2}q\{\nabla[\mathbf{B} \cdot (\mathbf{r'} \times \mathbf{v})] + \mathbf{r'} \times [(\mathbf{v} \cdot \nabla) \times \mathbf{B}) - (\mathbf{v} \cdot \nabla)(\mathbf{r'} \times \mathbf{B})\}$$
$$= \sum \frac{1}{2}q\{\nabla[\mathbf{B} \cdot (\mathbf{r'} \times \mathbf{v})]\}$$
$$= \nabla(\mathbf{m} \cdot \mathbf{B})$$

where $\mathbf{m} = \frac{1}{2}\sum_i q_i (\mathbf{r_i'} \times \mathbf{v_i'})$ is consistent with the formula $\mathbf{m} = \frac{1}{2}\int \mathbf{r'} \times \mathbf{J} d\mathbf{r'}$. The time average of (1.15) gives

$$\mathbf{F}_{m-dipole} = \frac{1}{2}\text{Re}\{\nabla[\mathbf{m} \cdot \mathbf{B}^*]\}, \tag{1.16}$$

which is the optical force for a magnetic dipole.

To the third order:

$$\sum q\{\frac{1}{6}\mathbf{r'r'r'}\vdots\nabla\nabla\nabla\mathbf{E}+\mathbf{v}\times(\frac{1}{2}\mathbf{r'r'}:\nabla\nabla\mathbf{B})\}$$

$$=\sum q\{\frac{1}{6}\mathbf{r'r'r'}\vdots\nabla\nabla\nabla\mathbf{E}-\mathbf{r'}\times(\dot{\mathbf{r'}}\mathbf{r'}:\nabla\nabla\mathbf{B}+\frac{1}{2}\mathbf{r'r'}:\nabla\nabla\dot{\mathbf{B}})\}$$

$$=\sum q\{\frac{1}{6}\mathbf{r'r'r'}\vdots\nabla\nabla\nabla\mathbf{E}-\mathbf{r'}\times\dot{\mathbf{r'}}\mathbf{r'}:\nabla\nabla\mathbf{B}+\frac{1}{2}\mathbf{r'}\times[\mathbf{r'r'}:\nabla\nabla(\nabla\times\mathbf{E})]\}$$

$$=\sum q\{\frac{1}{6}\mathbf{r'r'r'}\vdots\nabla\nabla\nabla\mathbf{E}-\mathbf{r'}\times\dot{\mathbf{r'}}\mathbf{r'}:\nabla\nabla\mathbf{B}+\frac{1}{2}\nabla[\mathbf{r'r'r'}\vdots\nabla\nabla\mathbf{E}]-\frac{1}{2}\mathbf{r'r'r'}\vdots\nabla\nabla\nabla\mathbf{E}\}$$

$$=\sum q\{\frac{1}{2}\nabla[\mathbf{r'r'r'}\vdots\nabla\nabla\mathbf{E}]-\frac{1}{3}\mathbf{r'r'r'}\vdots\nabla\nabla\nabla\mathbf{E}-\mathbf{r'}\times\dot{\mathbf{r'}}\mathbf{r'}:\nabla\nabla\mathbf{B}\}$$

$$=\sum q\{\frac{1}{6}\nabla[(\mathbf{r'r'r'}-\frac{3}{5}r'^2 r'_{[i}\delta_{jk]}\hat{e}_i\hat{e}_j\hat{e}_k)\vdots\nabla\nabla\mathbf{E}]+\frac{1}{3}\nabla[\mathbf{r'r'r'}\vdots\nabla\nabla\mathbf{E}]$$
$$+\frac{1}{10}\nabla[r'^2\mathbf{r'}\cdot\nabla^2\mathbf{E}]-\frac{1}{3}\mathbf{r'r'r'}\vdots\nabla\nabla\nabla\mathbf{E}-\mathbf{r'}\times\dot{\mathbf{r'}}\mathbf{r'}:\nabla\nabla\mathbf{B}\}$$

(1.17)

$$=\frac{\nabla}{6}[\vec{\mathbf{\Omega}}\vdots\nabla\nabla\mathbf{E}] \tag{1.18}$$

$$+\sum q\{\frac{1}{3}\nabla[\mathbf{r'r'r'}\vdots\nabla\nabla\mathbf{E}]+\frac{1}{10}\nabla[r'^2\mathbf{r'}\cdot\nabla^2\mathbf{E}]-\frac{1}{3}\mathbf{r'r'r'}\vdots\nabla\nabla\nabla\mathbf{E}-\mathbf{r'}\times\dot{\mathbf{r'}}\mathbf{r'}:\nabla\nabla\mathbf{B}\}$$

where

$$\vec{\mathbf{\Omega}}=\sum q\{\mathbf{r'r'r'}-\frac{3}{5}r'^2 r'_{[i}\delta_{jk]}\hat{e}_i\hat{e}_j\hat{e}_k\} \tag{1.19}$$

is the traceless octupole moment. The time average of (1.18) is

$$\mathbf{F}_{e-octopole}=\frac{\nabla}{12}\mathrm{Re}[\vec{\mathbf{\Omega}}:\nabla\nabla\mathbf{E}^*(\mathbf{r})], \tag{1.20}$$

which is the force associated with electric octupole. The second line of (1.18) is

$$\sum q\{\frac{1}{10}\nabla[r'^2\mathbf{r'}\cdot\nabla^2\mathbf{E}]+\frac{1}{3}\nabla[\mathbf{r'r'r'}\vdots\nabla\nabla\mathbf{E}]-\frac{1}{3}\mathbf{r'r'r'}\vdots\nabla\nabla\nabla\mathbf{E}-\mathbf{r'}\times\dot{\mathbf{r'}}\mathbf{r'}:\nabla\nabla\mathbf{B}\}$$

$$=q\{-\frac{\omega^2}{10c^2}\nabla[r'^2\mathbf{r'}\cdot\mathbf{E}]+\sum q\{\frac{1}{3}\nabla[\mathbf{r'r'r'}\vdots\nabla\nabla\mathbf{E}]-\frac{1}{3}[r_i'r_j'r_m'\nabla_i\nabla_m(\nabla_k E_j-\varepsilon_{jkl}\dot{B}_l)\hat{e}_k]-\mathbf{r'}\times\dot{\mathbf{r'}}\mathbf{r'}:\nabla\nabla\mathbf{B}\}$$

(1.21)

$$=-\frac{i\omega}{c}\nabla\{\vec{\mathbf{T}}\cdot\mathbf{E}\}+\sum q\{\frac{1}{3}\nabla[\mathbf{r'r'r'}\vdots\nabla\nabla\mathbf{E}]-\frac{1}{3}[r_i'r_j'r_m'\nabla_i\nabla_m(\nabla_k E_j-\varepsilon_{jkl}\dot{B}_l)\hat{e}_k]-\mathbf{r'}\times\dot{\mathbf{r'}}\mathbf{r'}:\nabla\nabla\mathbf{B}\}$$

(1.22)

where

$$\vec{\mathbf{T}} = \frac{1}{10}[\mathbf{r}'(\mathbf{r}'\cdot\mathbf{v}) - 2r'^2\,\mathbf{v}\mathbf{I}] = \frac{1}{10}\int\left[\mathbf{r}'(\mathbf{r}'\cdot\mathbf{J}) - 2r'^2\,\mathbf{J}\right]d\mathbf{r}' \quad (1.23)$$

is the toroidal dipole moment. The time average of the first term in (1.22) is

$$\mathbf{F}_{toroid} = -\frac{1}{2}\frac{\omega}{c^2}\nabla\,\mathrm{Im}\{\vec{\mathbf{T}}\cdot\mathbf{E}*\}, \quad (1.24)$$

which is the force associated with the Toroidal dipole moment. The remaining terms in (1.22) equals

$$= \sum q\{\frac{1}{3}\nabla[\mathbf{r}'\mathbf{r}'\mathbf{r}'\vdots\nabla\nabla\mathbf{E}] - \frac{1}{3}\mathbf{r}'\mathbf{r}'\mathbf{r}'\vdots\nabla\nabla\nabla\mathbf{E} - \mathbf{r}'\times\dot{\mathbf{r}}'\mathbf{r}':\nabla\nabla\mathbf{B}\}$$

$$= \sum q\{\frac{1}{3}[r_i'r_j'r_m'\nabla_i\nabla_m\varepsilon_{jkl}\dot{B}_l\hat{e}_k] - \mathbf{r}'\times\dot{\mathbf{r}}'\mathbf{r}':\nabla\nabla\mathbf{B}\}$$

$$= \sum q\{-\frac{1}{3}[2\dot{r}_i'r_j'r_m'\nabla_i\nabla_m\varepsilon_{jkl}B_l\hat{e}_k + r_i'\dot{r}_j'r_m'\nabla_i\nabla_m\varepsilon_{jkl}B_l\hat{e}_k] - \mathbf{r}'\times\dot{\mathbf{r}}'\mathbf{r}':\nabla\nabla\mathbf{B}\}$$

$$= \sum q\{\frac{2}{3}(\mathbf{v}\cdot\nabla)(\mathbf{r}'\cdot\nabla)(\mathbf{r}'\times\mathbf{B}) + \frac{1}{3}(\mathbf{r}'\cdot\nabla)(\mathbf{r}'\cdot\nabla)(\mathbf{v}\times\mathbf{B}) - \mathbf{r}'\times\dot{\mathbf{r}}'\mathbf{r}':\nabla\nabla\mathbf{B}\}$$

$$= \sum q\{\frac{1}{3}(\mathbf{r}'\cdot\nabla)(\mathbf{r}'\cdot\nabla)(\mathbf{v}\times\mathbf{B}) - \frac{1}{3}(\mathbf{v}\cdot\nabla)(\mathbf{r}'\cdot\nabla)(\mathbf{r}'\times\mathbf{B})\}$$

$$= \frac{\nabla}{3}[(\vec{\mathbf{Q}}_m:\nabla\mathbf{B})]$$

(1.25)

where

$$\ddot{\mathbf{Q}}_m = \sum q\{\frac{1}{2}[(\mathbf{r}'\times\mathbf{v})\mathbf{r}' + \mathbf{r}'(\mathbf{r}'\times\mathbf{v})] - \frac{1}{3}(\mathbf{r}'\times\mathbf{v})\cdot\mathbf{r}'\}$$
$$= \frac{1}{2}\left[\int(\mathbf{r}'\times\mathbf{J})\mathbf{r}'d\mathbf{r}' + \int\mathbf{r}'(\mathbf{r}'\times\mathbf{J})d\mathbf{r}'\right] - \frac{1}{3}trace\left[\int\mathbf{r}'(\mathbf{r}'\times\mathbf{J})d\mathbf{r}'\right] \quad (1.26)$$

is the traceless magnetic quadrupole moment. The time average of (1.25) is

$$\frac{\nabla}{6}\mathrm{Re}\{\ddot{\mathbf{Q}}_m:\nabla\mathbf{B}*\}, \quad (1.27)$$

which is the force associated with the magnetic quadrupole moment.

To summarize, the total force excluding the recoil force is

$$F_{total-radiation} = \frac{1}{2}\mathrm{Re}\{\nabla(\mathbf{p}\cdot\mathbf{E}*)\} + \frac{1}{2}\mathrm{Re}\{\nabla[\mathbf{m}\cdot\mathbf{B}*]\} + \frac{1}{12}\mathrm{Re}\{\nabla[\ddot{\mathbf{Q}}:\nabla\mathbf{E}*]\}$$
$$+ \frac{\nabla}{6}\mathrm{Re}\{\ddot{\mathbf{Q}}_m:\nabla\mathbf{B}*\} + \frac{\nabla}{12}\mathrm{Re}\{\ddot{\mathbf{\Omega}}:\nabla\nabla\mathbf{E}*(\mathbf{r})\} - \frac{1}{2}\frac{\omega}{c^2}\nabla\,\mathrm{Im}\{\vec{\mathbf{T}}\cdot\mathbf{E}*\} \quad (1.28)$$

*Appendix A: Derivation based on the optical force acting on an electric dipole*

We shall show that the time averaged optical force acting on a small particle can be expanded in a multipole expansion. A particle illuminated by an external incident field is exposed to two types of field, namely the initial incident field and the field produced by the multipoles. As a result, the particle also experiences two types of optical forces, namely the force induced directly by the incident field (Type 1 forces) and the force induced by the multipoles (Type 2 forces). We shall derive the former force first, and then derive the latter force.

In the main text, we derive the time averaged Type 1 optical force acting on the multipoles by starting from the scratch, i.e. by considering a collection of point charges that build up the electrodynamics multipoles, one can derive the optical force from the Lorentz force. Here, we shall take an alternative approach. We shall start from the formula for the electric dipole force, and then evaluate the optical forces acting on the higher order multipoles through a limiting procedure.

Consider a composite particle consists of two nearby dipoles having total dipole moment $\mathbf{p}_t = \mathbf{p}_1 + \mathbf{p}_2$, and situated in an external harmonic field $\mathbf{E}$. Here we shall denote this composite particle as a C4-particle, because it is composed of four charges. Such C4-particle possesses an electric dipole moment, a magnetic dipole moment, and an electric quadrupole moment. We know that the time averaged optical force acting on a single dipole is given by

$$\mathbf{F}_{e-dipole} = \frac{1}{2}\text{Re}\left\{\nabla[\mathbf{p}\cdot\mathbf{E}^*]\right\}, \qquad (1)$$

where $\mathbf{p}$ is the dipole moment and $\mathbf{E}$ is the external incident field, and the gradient operator does not operate on $\mathbf{p}$, and this is actually the first order force. Accordingly, we use this to calculate the optical force acting on the C4-particle, which is

$$\mathbf{F} = \frac{1}{2}\text{Re}\left\{\nabla[\mathbf{p_1}\cdot\mathbf{E}^*(\mathbf{r}+\mathbf{d}/2)]\right\} + \frac{1}{2}\text{Re}\left\{\nabla[\mathbf{p}_2\cdot\mathbf{E}^*(\mathbf{r}-\mathbf{d}/2)]\right\}, \qquad (2)$$

where $\bar{d}$ is the separation between the two composing dipoles. Since the separation

for the pair of dipole is small, we also take the first order to get

$$\mathbf{F} \simeq \frac{\nabla}{2}\mathrm{Re}\left\{\mathbf{p_1}\cdot\left[\mathbf{E}^*(\mathbf{r})+\frac{\mathbf{d}\cdot\nabla}{2}\mathbf{E}^*(\mathbf{r})\right]\right\}+\frac{\nabla}{2}\mathrm{Re}\left\{\mathbf{p_2}\cdot\left[\mathbf{E}^*(\mathbf{r})-\frac{\mathbf{d}\cdot\nabla}{2}\mathbf{E}^*(\mathbf{r})\right]\right\}$$
$$=\frac{\nabla}{2}\mathrm{Re}\left\{\left[(\mathbf{p_1}+\mathbf{p_2})\cdot\mathbf{E}^*(\mathbf{r})\right]+\frac{\mathbf{p_1}-\mathbf{p_2}}{2}\cdot(\mathbf{d}\cdot\nabla)\mathbf{E}^*(\mathbf{r})\right\} \quad (3)$$

The first term in the second line of (3),

$$\mathbf{F}_{e-dipole}=\frac{1}{2}\mathrm{Re}\left\{\nabla\left(\mathbf{p_t}\cdot\mathbf{E}^*\right)\right\}, \quad (4)$$

is the optical force that the incident field directly acts on the total electric dipole moment of the particle, where the gradient operator does not operate on $\mathbf{p}_t$. Now consider the second term in the second line of (3). Let

$$\mathbf{p}_r=\frac{1}{2}(\mathbf{p_1}-\mathbf{p_2}), \quad (5)$$

so we have

$$\mathbf{F}=\mathbf{F}_{e-dipole}+\frac{1}{2}\nabla\mathrm{Re}\left\{\mathbf{p}_r\cdot(\mathbf{d}\cdot\nabla)\mathbf{E}^*(\mathbf{r})\right\}$$
$$=\mathbf{F}_{e-dipole}+\frac{1}{2}\nabla\mathrm{Re}\left\{\mathbf{dp}_r:\nabla\mathbf{E}^*(\mathbf{r})\right\} \quad (6)$$
$$=\mathbf{F}_{e-dipole}+\frac{\nabla\mathrm{Re}\left\{(\mathbf{dp}_r+\mathbf{p}_r\mathbf{d}):\nabla\mathbf{E}^*(\mathbf{r})\right\}}{4}+\frac{\nabla\mathrm{Re}\left\{(\mathbf{dp}_r-\mathbf{p}_r\mathbf{d}):\nabla\mathbf{E}^*(\mathbf{r})\right\}}{4}$$

We get

$$\mathbf{F}=\mathbf{F}_{e-dipole}+\mathbf{F}_{e-quadrupole}+\frac{\nabla}{4}\mathrm{Re}\left\{(\mathbf{dp}_r-\mathbf{p}_r\mathbf{d}):\nabla\mathbf{E}^*(\mathbf{r})\right\}, \quad (7)$$

where

$$\mathbf{F}_{e-quadrupole}=\frac{\nabla}{12}\mathrm{Re}\left\{\ddot{\mathbf{Q}}:\nabla\mathbf{E}^*(\mathbf{r})\right\}, \quad (8)$$

and the electric quadrupole moment $\ddot{\mathbf{Q}}$ can be defined as

$$3\sum_i q_i\mathbf{r_i'}\mathbf{r_i'}, \quad (2.9)$$

since

$$\mathbf{r}_1=\left(\frac{\mathbf{d}}{2}+\frac{\mathbf{s_1}}{2}\right), \quad \mathbf{r}_2=\left(\frac{\mathbf{d}}{2}-\frac{\mathbf{s_1}}{2}\right), \quad \mathbf{r}_3=\left(-\frac{\mathbf{d}}{2}+\frac{\mathbf{s_1}}{2}\right), \quad \mathbf{r}_4=\left(-\frac{\mathbf{d}}{2}-\frac{\mathbf{s_1}}{2}\right), \quad (2.10)$$

and

$$\vec{Q} = 3\sum_i q_i \mathbf{r_i'}\mathbf{r_i'} = \frac{3}{2}[q_1(\mathbf{ds_1} + \mathbf{s_1 d}) - q_2(\mathbf{ds_2} + \mathbf{s_2 d})] = 3(\mathbf{dp_r} + \mathbf{p_r d}) \qquad (2.11)$$

However, since $\mathbf{I}:\nabla \mathbf{E} = \nabla \cdot \mathbf{E} = 0$, we can de-trace $\vec{Q}$ without affecting (8):

$$\vec{Q} = \sum_i q_i(3\mathbf{r_i'}\mathbf{r_i'} - r_i'^2 \mathbf{I}), \qquad (2.12)$$

which is consistent with the formula $\vec{Q} = \sum_i q_i(3\mathbf{r_i'}\mathbf{r_i'} - r_i'^2 \mathbf{I}) = \int \rho(3\mathbf{r'r'} - r'^2 \mathbf{I})d\mathbf{r'}$ where $\rho = \sum_i q_i \delta(\mathbf{r} - \mathbf{r}_i)$, and $\mathbf{r}_i$ is the position vector of the charge.

It can be shown that the last term in (7) can be simplified into:

$$\frac{1}{4}\text{Re}\{\nabla[(\mathbf{d}\times\mathbf{p_r})\cdot(\nabla\times\mathbf{E}^*)]\}. \qquad (13)$$

Applying the Maxwell equation $\nabla\times\mathbf{E} = -\frac{\partial \mathbf{B}}{\partial t} = i\omega\mathbf{B}$ to (13), one obtains

$$-\frac{1}{4}\text{Re}\{\nabla[(\mathbf{d}\times\mathbf{p_r})\cdot i\omega\mathbf{B}^*]\} = \frac{1}{4}\text{Re}\{\nabla[(\mathbf{d}\times\dot{\mathbf{p}}_\mathbf{r})\cdot\mathbf{B}^*]\}$$
$$= \frac{1}{2}\text{Re}\{\nabla[\mathbf{m}\cdot\mathbf{B}^*]\} = \mathbf{F}_{m-dipole} \qquad (14)$$

where we have used the fact that we calculate only the time-average force, and the magnetic dipole moment is given by

$$\mathbf{m} = \frac{1}{2}\sum_i q_i(\mathbf{r_i'}\times\mathbf{v_i'}) = \frac{1}{2}[q_1(\frac{\mathbf{d}}{2}+\frac{\mathbf{s_1}}{2})\times(\frac{\dot{\mathbf{s}}_1}{2}) - q_1(\frac{\mathbf{d}}{2}-\frac{\mathbf{s_1}}{2})\times(-\frac{\dot{\mathbf{s}}_1}{2})$$
$$+ q_2(-\frac{\mathbf{d}}{2}+\frac{\mathbf{s_2}}{2})\times(\frac{\dot{\mathbf{s}}_2}{2}) - q_2(-\frac{\mathbf{d}}{2}-\frac{\mathbf{s_2}}{2})\times(-\frac{\dot{\mathbf{s}}_2}{2})]$$
$$= \frac{1}{4}[\mathbf{d}\times\dot{\mathbf{p}}_1 - \mathbf{d}\times\dot{\mathbf{p}}_2] = \frac{1}{2}[\mathbf{d}\times\dot{\mathbf{p}}_\mathbf{r}]$$

which is consistent with the formula $\mathbf{m} = \frac{1}{2}\int \mathbf{r'}\times\mathbf{J}d\mathbf{r'}$. We should note here that the charge can only move around its equilibrium position, so $\mathbf{v_i} = \dot{\mathbf{s}}_\mathbf{i} = -i\omega\mathbf{s_i}$

(15)

As a result, for the C4-particle which possesses an electric dipole moment, a magnetic dipole moment, and an electric quadrupole moment, the force that the incident field directly acts on the particle is given by

$$\mathbf{F} = \frac{1}{2}\operatorname{Re}\left\{\nabla[\mathbf{p}\cdot\mathbf{E}^*]\right\} + \frac{1}{2}\operatorname{Re}\left\{\nabla[\mathbf{m}\cdot\mathbf{B}^*]\right\} + \frac{1}{12}\operatorname{Re}\left\{\nabla\left[\ddot{\mathbf{Q}}:\nabla\mathbf{E}^*\right]\right\}. \tag{16}$$

While the Type 1 forces are given by (16), one still has to determine the Type 2 forces. This can be calculated by using the Maxwell stress tensor. Consider, for example, in order to evaluate the interference induced optical force between the electric dipole and magnetic dipole, one can integrate the Maxwell stress tensor with the field only given by the electric and magnetic dipole. This amounts to the interaction between the electric dipole and the magnetic dipole. The procedure for the other multipoles is similar. With some straight forward but tedious calculation, one finally arrives at the total optical force:

$$\begin{aligned}\mathbf{F}_{total} = &\frac{1}{2}\operatorname{Re}\left\{\nabla[\mathbf{p}\cdot\mathbf{E}^*]\right\} + \frac{1}{2}\operatorname{Re}\left\{\nabla[\mathbf{m}\cdot\mathbf{B}^*]\right\} + \frac{1}{12}\operatorname{Re}\left\{\nabla\left[\ddot{\mathbf{Q}}:\nabla\mathbf{E}^*\right]\right\} \\ &- \frac{k^3\omega\mu_0}{12\pi}\operatorname{Re}\left\{\mathbf{p}\times\mathbf{m}^*\right\} - \frac{ck^5 Z_0}{120\pi}\operatorname{Im}\left\{\ddot{Q}\cdot\mathbf{P}^*\right\}\end{aligned}. \tag{17}$$

We note that the optical force has the same form irrespective to the origin of the multipole moments, be it due to point charges, magnetic charges, dipoles, or else. Thus, (17) truly represents the optical force acting on a particle with electric dipole moment $\mathbf{p}$, magnetic dipole moment $\mathbf{m}$, and electric quadrupole moment $\ddot{\mathbf{Q}}$, irrespective to the origin of the moments.

*Appendix B: Derivation of the optical force acting on a small particle characterized by multipole moments up to electric octopole*

To derive the expression of the higher order multipole force, one could start everything from scratch by considering a collection of charges that builds up the multipoles, just as in the main text. The extinction force can then be derived from the Lorentz force after proper limiting procedure. Alternatively, one could follow the approach of

Appendix A. By building higher order multipole from dipoles, one can then evaluate the force for the higher order multipole from the dipolar force expression. The latter approach is somewhat simpler. Here we take an even simpler approach. We consider a composite particle (denoted by C8-particle here, as it is composed of eight charges) consists of a pair of C4-particle discussed in Appendix A, except that for simplicity, we take $\mathbf{p}_1 = -\mathbf{p}_2 = \mathbf{p}$ for the first C4-particle, and we take $\mathbf{p}_1 = -\mathbf{p}_2 = -\mathbf{p}$ for the second C4-particle. These C4-particles possess electric quadrupole moments and magnetic dipole moments, but not net dipole moments as discussed in Appendix A. According to the second line of (6), the optical force acting on the composite C8-particle is given by

$$\mathbf{F}_{C8} = \nabla \operatorname{Re}\{\mathbf{dp} : [\nabla \mathbf{E}^*(\mathbf{r}+\mathbf{l}/2) - \nabla \mathbf{E}^*(\mathbf{r}-\mathbf{l}/2)]\}/2, \qquad (18)$$

where $\mathbf{l}$ is the separation between the centers of the two C4-particles. Invoking a Taylor series expansion for $\mathbf{E}^*$, one arrives at

$$\begin{aligned}
\mathbf{F}_{C8} &= (\nabla/2)\operatorname{Re}\{\mathbf{dp} : \nabla(\mathbf{l}\cdot\nabla)\mathbf{E}^*(\mathbf{r})\} \\
&= (\nabla/2)\operatorname{Re}\{\mathbf{ldp} : \nabla\nabla\mathbf{E}^*(\mathbf{r})\} \\
&= \frac{\nabla}{2}\operatorname{Re}\{Sym(\mathbf{ldp}) : \nabla\nabla\mathbf{E}^*(\mathbf{r})\} + \frac{\nabla}{2}\operatorname{Re}\{[\mathbf{ldp} - Sym(\mathbf{ldp})] : \nabla\nabla\mathbf{E}^*(\mathbf{r})\}, \quad (19) \\
&= \frac{\nabla}{12}\operatorname{Re}\{\ddot{\Omega} : \nabla\nabla\mathbf{E}^*(\mathbf{r})\} + \frac{\nabla}{12}\operatorname{Re}\{[6 Sym(\mathbf{ldp}) - \ddot{\Omega}] : \nabla\nabla\mathbf{E}^*(\mathbf{r})\} \\
&\quad + \frac{\nabla}{2}\operatorname{Re}\{[\mathbf{ldp} - Sym(\mathbf{ldp})] : \nabla\nabla\mathbf{E}^*(\mathbf{r})\}
\end{aligned}$$

where

$$\begin{aligned}
\ddot{\Omega} &= 6\{Sym(\mathbf{ldp}) - \frac{1}{5}[Sym(\mathbf{ldp})_{ill}\delta_{jk} + Sym(\mathbf{ldp})_{jll}\delta_{ik} + Sym(\mathbf{ldp})_{kll}\delta_{ij}]\hat{e}_i\hat{e}_j\hat{e}_k\} \\
&= \int \rho \mathbf{r'r'r'}d\mathbf{r'} - \frac{3}{5}\int \rho r'^2 \mathbf{r'}d\mathbf{r'}
\end{aligned} \qquad (20)$$

is the symmetric traceless electric octopole moment, and

$$Sym(\mathbf{ldp}) = \frac{\mathbf{ldp} + \mathbf{lpd} + \mathbf{pld} + \mathbf{pdl} + \mathbf{dpl} + \mathbf{dlp}}{6} \qquad (21)$$

Is all permutation of $\mathbf{l}$, $\mathbf{d}$, and $\mathbf{p}$ divide by six. We note that unlike in electrostatic, in electrodynamics, the trace of the octopole moment does not vanish. Consequently, we have an additional term in the middle of the final line of (19). Next, consider the

last term in the last line of (19). Noting that

$$\begin{aligned}
(\mathbf{ldp \cdot ldp}) &: \nabla\nabla \mathbf{E}^* = 0 \\
(\mathbf{ldp \cdot lpd}) &: \nabla\nabla \mathbf{E}^* = \mathbf{l}(\mathbf{d}\times\mathbf{p}) : \nabla\nabla\times\mathbf{E}^* \\
(\mathbf{ldp \cdot pld}) &: \nabla\nabla \mathbf{E}^* = \mathbf{l}(\mathbf{d}\times\mathbf{p}) : \nabla\nabla\times\mathbf{E}^* \\
(\mathbf{ldp \cdot pdl}) &: \nabla\nabla \mathbf{E}^* = \mathbf{d}(\mathbf{l}\times\mathbf{p}) : \nabla\nabla\times\mathbf{E}^* \\
(\mathbf{ldp \cdot dpl}) &: \nabla\nabla \mathbf{E}^* = \mathbf{d}(\mathbf{l}\times\mathbf{p}) : \nabla\nabla\times\mathbf{E}^* \\
(\mathbf{ldp \cdot dlp}) &: \nabla\nabla \mathbf{E}^* = 0
\end{aligned} \qquad (22)$$

The last term of the last line of (19) can be written as

$$\frac{\nabla}{2}\operatorname{Re}\left\{\frac{\mathbf{l}(\mathbf{d}\times\mathbf{p})+\mathbf{d}(\mathbf{l}\times\mathbf{p})}{3} : \nabla(-i\omega\mathbf{B}^*)\right\}, \qquad (23)$$

where we have made use of the Maxwell equation $\nabla\times\mathbf{E} = i\omega\mathbf{B}$. One can split $\mathbf{l}(\mathbf{d}\times\mathbf{p})+\mathbf{d}(\mathbf{l}\times\mathbf{p})$ into two equal half, and then exchange the subscript $i$ and $j$ of the second part:

$$\frac{\nabla}{2}\operatorname{Re}\left\{-i\omega\frac{l_i(\mathbf{d}\times\mathbf{p})_j+d_i(\mathbf{l}\times\mathbf{p})_j}{6} : \nabla_i B_j^* - i\omega\frac{(\mathbf{d}\times\mathbf{p})_i l_j+(\mathbf{l}\times\mathbf{p})_i d_j}{6} : \nabla_j B_j^*\right\}. \qquad (24)$$

Using the Maxwell equation

$$\nabla_j B_i^* = \nabla_i B_j^* - \varepsilon_{ijk}\frac{i\omega}{c^2}E_k^*, \qquad (25)$$

(24) becomes

$$\frac{\nabla}{2}\text{Re}\left\{\begin{array}{l}-i\omega\dfrac{l_i(\mathbf{d}\times\mathbf{p})_j+d_i(\mathbf{l}\times\mathbf{p})_j}{6}:\nabla_iB_j{}^*\\-i\omega\dfrac{(\mathbf{d}\times\mathbf{p})_il_j+(\mathbf{l}\times\mathbf{p})_id_j}{6}:\left(\nabla_iB_j{}^*-\sum_k\varepsilon_{ijk}\dfrac{i\omega}{c^2}E_k{}^*\right)\end{array}\right\}$$

$$=\frac{\nabla}{12}\text{Re}\{[\mathbf{l}(\mathbf{d}\times\mathbf{p})+\mathbf{d}(\mathbf{l}\times\mathbf{p})+(\mathbf{d}\times\mathbf{p})\mathbf{l}+(\mathbf{l}\times\mathbf{p})\mathbf{d}]:[-i\omega\nabla\mathbf{B}^*)]\}$$

$$-\frac{\nabla}{12}\frac{\omega^2}{c^2}\text{Re}\{[(\mathbf{d}\times\mathbf{p})\times\mathbf{l}+(\mathbf{l}\times\mathbf{p})\times\mathbf{d}]\cdot\mathbf{E}^*\}$$

$$=-\frac{\nabla}{12}\text{Re}\{-i\omega[\mathbf{l}(\mathbf{d}\times\mathbf{p})+\mathbf{d}(\mathbf{l}\times\mathbf{p})+(\mathbf{d}\times\mathbf{p})\mathbf{l}+(\mathbf{l}\times\mathbf{p})\mathbf{d}]:\nabla\mathbf{B}^*\}$$

$$-\frac{\nabla}{12}\frac{\omega^2}{c^2}\text{Re}\{[(\mathbf{d}\times\mathbf{p})\times\mathbf{l}+(\mathbf{l}\times\mathbf{p})\times\mathbf{d}]\cdot\mathbf{E}^*\}$$

$$=-\frac{\nabla}{12}\text{Re}\{[\mathbf{l}(\mathbf{d}\times\dot{\mathbf{p}})+\mathbf{d}(\mathbf{l}\times\dot{\mathbf{p}})+(\mathbf{d}\times\dot{\mathbf{p}})\mathbf{l}+(\mathbf{l}\times\dot{\mathbf{p}})\mathbf{d}]:\nabla\mathbf{B}^*\}$$

$$-\frac{\nabla}{12}\frac{\omega^2}{c^2}\text{Re}\{[(\mathbf{d}\times\mathbf{p})\times\mathbf{l}+(\mathbf{l}\times\mathbf{p})\times\mathbf{d}]\cdot\mathbf{E}^*\}$$

$$=\frac{\nabla}{6}\text{Re}\{\ddot{\mathbf{Q}}_m:\nabla\mathbf{B}^*\}-\frac{\nabla}{12}\frac{\omega^2}{c^2}\text{Re}\{[(\mathbf{d}\times\mathbf{p})\times\mathbf{l}+(\mathbf{l}\times\mathbf{p})\times\mathbf{d}]\cdot\mathbf{E}^*\} \qquad ,(26)$$

where

$$\ddot{\mathbf{Q}}_\mathbf{m}=\frac{1}{2}[\mathbf{l}(\mathbf{d}\times\dot{\mathbf{p}})+\mathbf{d}(\mathbf{l}\times\dot{\mathbf{p}})+(\mathbf{d}\times\dot{\mathbf{p}})\mathbf{l}+(\mathbf{l}\times\dot{\mathbf{p}})\mathbf{d}]$$
$$=\frac{1}{2}\left[\int(\mathbf{r'}\times\mathbf{J})\mathbf{r'}d\mathbf{r'}+\int\mathbf{r'}(\mathbf{r'}\times\mathbf{J})d\mathbf{r'}\right]-\frac{1}{3}trace\left[\int\mathbf{r'}(\mathbf{r'}\times\mathbf{J})d\mathbf{r'}\right] \qquad (27)$$

is the symmetric traceless magnetic quadrupole moment. Accordingly, (19) can be further simplified into

$$\mathbf{F}_{C8}=\frac{\nabla}{12}\text{Re}\{\ddot{\mathbf{\Omega}}:\nabla\nabla\mathbf{E}^*(\mathbf{r})\}+\frac{\nabla}{6}\text{Re}\{\ddot{\mathbf{Q}}_m:\nabla\mathbf{B}^*\}$$
$$+\frac{1}{10}\nabla\text{Re}\{Sym(\mathbf{ldp})_{kll}\nabla_i\nabla_i\mathbf{E}_k{}^*(\mathbf{r})\}-\frac{\nabla}{12}\frac{\omega^2}{c^2}\text{Re}\{[(\mathbf{d}\times\mathbf{p})\times\mathbf{l}+(\mathbf{l}\times\mathbf{p})\times\mathbf{d}]\cdot\mathbf{E}^*\} \qquad ,(28)$$

The last two term in (28) can be combined to give

$$\frac{1}{10}\nabla\text{Re}\left\{Sym(\mathbf{ldp})_{kll}\nabla^2 E_k*(\mathbf{r})\right\} - \frac{\nabla}{12}\frac{\omega^2}{c^2}\text{Re}\left\{\left[(\mathbf{d}\times\mathbf{p})\times\mathbf{l} + (\mathbf{l}\times\mathbf{p})\times\mathbf{d}\right]\cdot\mathbf{E}*\right\}$$

$$= -\frac{1}{10}\frac{\omega^2}{c^2}\nabla\text{Re}\left\{Sym(\mathbf{ldp})_{kll}E_k*(\mathbf{r})\right\} - \frac{\nabla}{12}\frac{\omega^2}{c^2}\text{Re}\left\{\left[(\mathbf{d}\times\mathbf{p})\times\mathbf{l} + (\mathbf{l}\times\mathbf{p})\times\mathbf{d}\right]\cdot\mathbf{E}*\right\}$$

$$= -\frac{1}{30}\frac{\omega^2}{c^2}\nabla\text{Re}\left\{\left[(\mathbf{d}\cdot\mathbf{p})\mathbf{l} + (\mathbf{d}\cdot\mathbf{l})\mathbf{p} + (\mathbf{l}\cdot\mathbf{p})\mathbf{d}\right]\cdot\mathbf{E}*\right\} + \frac{\nabla}{12}\frac{\omega^2}{c^2}\text{Re}\left\{\left[(\mathbf{l}\cdot\mathbf{p})\mathbf{d} - (\mathbf{d}\cdot\mathbf{l})\mathbf{p} + (\mathbf{d}\cdot\mathbf{p})\mathbf{l} - (\mathbf{d}\cdot\mathbf{l})\mathbf{p}\right]\cdot\mathbf{E}*\right\}$$

$$= -\frac{1}{20}\frac{\omega^2}{c^2}\nabla\text{Re}\left\{-(\mathbf{l}\cdot\mathbf{p})\mathbf{d} - (\mathbf{d}\cdot\mathbf{p})\mathbf{l} + 4(\mathbf{d}\cdot\mathbf{l})\mathbf{p}\right\}\cdot\mathbf{E}*$$

$$= -\frac{1}{2}\frac{\omega}{c^2}\nabla\text{Im}\left\{\mathbf{T}\cdot\mathbf{E}*\right\}$$

(29)

where

$$\mathbf{T} = \tfrac{1}{10}\int\left[\mathbf{r}'(\mathbf{r}'\cdot\mathbf{J}) - 2r'^2\mathbf{J}\right]d\mathbf{r}'$$
$$= \left[-(\mathbf{l}\cdot\dot{\mathbf{p}})\mathbf{d} - (\mathbf{d}\cdot\dot{\mathbf{p}})\mathbf{l} + 4(\mathbf{d}\cdot\mathbf{l})\dot{\mathbf{p}}\right]/10$$

(30)

is the toroidal dipole moment.

Consequently, the Type 1 forces acting on the C8 particle is

$$\mathbf{F}_{C8} = -\frac{\nabla}{12}\text{Re}\left\{\ddot{\mathbf{\Omega}}:\nabla\nabla\mathbf{E}*(\mathbf{r})\right\} + \frac{\nabla}{6}\text{Re}\left\{\ddot{\mathbf{Q}}_m:\nabla\mathbf{B}*\right\} - \frac{1}{2}\frac{\omega}{c^2}\nabla\text{Im}\left\{\mathbf{T}\cdot\mathbf{E}*\right\}.$$ (31)

Since the optical force has the same form irrespective to the origin of the moments, one can infer from (31) that the optical force acting on an electric octopole, a magnetic quadrupole, and a toroid dipole are given by the first, second, and third term of (31) respectively. Consequently, the total force acting on a particle possesses multipole moments up to electric octopole order is given by

$$\mathbf{F}_{total} = \frac{1}{2}\text{Re}\left\{\nabla\left(\mathbf{p}\cdot\mathbf{E}*\right)\right\} + \frac{1}{2}\text{Re}\left\{\nabla\left[\mathbf{m}\cdot\mathbf{B}*\right]\right\} + \frac{1}{12}\text{Re}\left\{\nabla\left[\ddot{\mathbf{Q}}:\nabla\mathbf{E}*\right]\right\}$$
$$+ \frac{\nabla}{6}\text{Re}\left\{\ddot{\mathbf{Q}}_m:\nabla\mathbf{B}*\right\} + \frac{\nabla}{12}\text{Re}\left\{\ddot{\mathbf{\Omega}}:\nabla\nabla\mathbf{E}*(\mathbf{r})\right\} - \frac{1}{2}\frac{\omega}{c^2}\nabla\text{Im}\left\{\mathbf{T}\cdot\mathbf{E}*\right\}$$
$$- \frac{k^3\omega\mu_0}{12\pi}\text{Re}\left\{\mathbf{p}\times\mathbf{m}*\right\} - \frac{ck^5 Z_0}{120\pi}\text{Im}\left\{\ddot{\mathbf{Q}}\cdot\mathbf{P}*\right\}$$

(32)

We note that we have ignored the interaction between the electric octopole, magnetic quadrupole, and toroid dipole with other multipole moments. The reason is that these terms are one order higher in $d$ than (32), and thus can be ignored.